\documentclass[aps,prb,twocolumn,amsmath,amssymb,nofootinbib,superscriptaddress,floatfix,eqsecnum,]{revtex4}

\usepackage{amsmath}
\usepackage{amssymb}
\usepackage{amsthm}
\usepackage[pdftex]{color} 
\usepackage{graphicx}
\usepackage{dcolumn} 
\usepackage{bm} 
\usepackage{longtable}
\usepackage{ulem}   
\newcommand{\bs}[1]{{\boldsymbol{#1}}}

\begin{document}

\title{
Ginzburg-Landau description of twin 
boundaries in noncentrosymmetric superconductors
      }

\author{Michael Achermann} 
\affiliation{
Institute for Theoretical Physics,
ETH Zurich, 8093 Zurich, Switzerland
            } 

\author{Titus Neupert} 
\affiliation{
Princeton Center for Theoretical Science, Princeton University, Princeton, New Jersey 08544, USA
            } 

\author{Emiko Arahata} 
\affiliation{
Department of Basic Science, The University of Tokyo,
3-8-1 Komaba, Meguro-ku, Tokyo, 153-8902, Japan
            } 

\author{Manfred Sigrist} 
\affiliation{
Institute for Theoretical Physics,
ETH Zurich, 8093 Zurich, Switzerland
            } 

\date{\today}

\begin{abstract}
We study theoretically a model for twin boundaries in superconductors with Rashba spin-orbit coupling, 
which can be relevant to both three-dimensional noncentrosymmetric tetragonal crystals and
two-dimensional gated superconductors such as the LaAlO$_3$/SrTiO$_3$ interface.
In both cases, the broken inversion symmetry allows for a coexistence of singlet and triplet
pairing. Within the framework of a Ginzburg-Landau theory, we 
identify two $\mathbb{Z}_2$ symmetries that are broken via two consecutive second order phase transitions
as the temperature is lowered.
We show that a time-reversal symmetry breaking superconducting state 
nucleates near the twin boundary if singlet and triplet pairing amplitudes are of comparable magnitude.
As a consequence, the tendency towards ferromagnetic order is locally increased
along with the emergence of spontaneous supercurrents parallel to the twin boundary. 
Spin currents, which are present in the form of Andreev bound states, 
are found enhanced in the time-reversal broken phases.
\end{abstract}

\maketitle


\medskip
\section{
Introduction
        }

Non-centrosymmetric superconductors are endowed with a number of peculiar properties that originate from
the modification of the spin structure of electronic states due to spin-orbit coupling. Among other features
the mixing of even and odd parity pairing channels is found, allowed by the lack of inversion 
symmetry~\cite{Gorkov01,Frigeri04,Yokoyama07,Sigrist09,Springer}. Recent interest in noncentrosymmetric superconductors
has been launched by the discovery of heavy Fermion superconductors in this class
such as CePt$_3$Si~\cite{BAU04}, CeRhSi$_3$~\cite{Kimura05} and CeIrSi$_3$\cite{Onuki06}, which
show superconductivity together with magnetic order~\cite{Springer}. This interplay of magnetism 
and superconductivity may give rise to unusual physical properties. 
Annother intriguing feature, the possible realization of edge states (Andreev bound states), has triggered many
theoretical studies, as they can arise as a consequence of the topological nature of the superconducting phase
in noncentrosymmetric superconductors and give rise of zero-bias anomalies in quasiparticle tunneling 
or non-trivial spin Hall response~\cite{Yokoyama05,Iniotakis07,Vorontsov08,Tanaka09,Sato06}.

Twin boundaries provide a particular environment displaying intriguing physical properties in 
noncentrosymmetric superconductors. Twin domain formation is probable to occur in noncentrosymmetric materials and arises due to degenerate realizations of broken inversion symmetry 
in crystals. Spin-orbit coupling specific to noncentrosymmetricity is different but symmetry-wise 
equivalent in these twin domains and leads to different mixed-parity pairing states. 
At twin boundaries these different pairing states have to be matched and may lead to new
superconducting phases. Among these may be phases that break time reversal symmetry
and can support fractionally quantized vortices~\cite{Iniotakis08}. Even local ferromagnetic order
and spontaneous supercurrents are possible~\cite{Arahata12}.

In this paper, we discuss features of noncentrosymmetric superconductivity at twin boundaries 
within a Ginzburg-Landau (GL) formulation. In the heavy Fermion 
superconductors mentioned above two types of twin domains give rise to Rashba-type spin-orbit
coupling of opposite sign. The GL approach allows us to analyze the symmetry properties of the
superconducting phase in the vicinity of the twin boundary. In a first step, we examine the
conditions for broken time reversal symmetry (TRS) as was discussed based on quasi-classical 
theory in Ref.~\onlinecite{Iniotakis08}. Interestingly, the TRS broken state does not show magnetic properties
unlike the generic TRS breaking superconducting states~\cite{Sigrist91b,Arahata12}. A second Ising-like
symmetry breaking transition leads to a superconducting twin boundary phase which develops both
a spin magnetization and a supercurrent. Our analysis shows that this secondary transition
can be viewed as a spontaneous spin Hall effect, as the twin boundary state, very much like the
edge states, carry a spin current.~\cite{Brydon13} With our GL discussion we confirm and extend the Bogolyubov-deGennes 
theory of Ref.~\onlinecite{Arahata12}.

The appearance of a finite spin magnetization relies on enhanced magnetic correlations. This may
be naturally true in the above heavy Fermion compounds, whose superconducting phase is closely linked to
a magnetic quantum phase transition. The two-dimensional superconductors at interfaces in heterostructures
such as the LaAlO$_3$/SrTiO$_3$ interface provide another possible environment for the discussed physics~\cite{Reyren07}.
There, the noncentrosymmetric structure of the interface induces a Rashba-like spin-orbit coupling. While it
may be difficult to introduce twin boundaries here, we may imagine that the spin-orbit coupling could
be spatially varied through artificial structuring, leading to boundaries between regions of different spin-orbit coupling
strengths. This may then lead to similar physics as at twin boundaries \cite{Aoyama12}. Interesting in this context is also
the fact that the LaAlO$_3$/SrTiO$_3$ interface shows some trend towards ferromagnetism which facilitates the
occurrence of a secondary magnetic transition~\cite{Brinkman07,Bert11,Li11}.  


\medskip
\section{
Microscopic mean-field description
        }
\label{sec: microscopics}

Before introducing the GL description 
we address some aspects of twin boundaries in a noncentrosymmetric superconductor 
from the point of view of a simplified Bogolyubov-deGennes formulation. We consider
first electrons in a three-dimensional tetragonal crystal 
[$d=3$, $\bs{x}=(x^{\ }_1,x^{\ }_2,x^{\ }_3)^{\mathsf{T}}$, which can easily be
reduced to the two-dimensional case of an interface with $d=2$, $\bs{x}=(x^{\ }_1,x^{\ }_2)^{\mathsf{T}}$].
Here, noncentrosymmetricity refers to the lack of symmetry under reflection 
at the basal plane of a tetragonal crystal. 
This is implemented in the Hamiltonian by introducing a Rashba-like spin-orbit coupling, leading to
a TRS spin splitting of the electronic states. On a tetragonal lattice (square lattice in 2D), 
a representative Hamiltonian 
including only nearest-neighbor hopping can be written as
\begin{subequations}
\begin{equation}
\begin{split}
H:=&\,\sum_{\bs{k}\in\text{BZ}}
\hat{c}^\dagger(\bs{k})\mathcal{H}(\bs{k})\hat{c}(\bs{k}),
\\
\mathcal{H}(\bs{k}):=&\,
\sum_{i=1}^d 2 t^{\ }_i\,\cos\, (a^{\ }_i k^{\ }_i)
+\alpha\, \bs{g}(\bs{k})\cdot \bs{\sigma}-\mu,
\label{non-interacting H}
\end{split}
\end{equation}
where $\mu$ is the chemical potential, 
$t^{\ }_1=t^{\ }_2$ and $t^{\ }_3$ are the nearest neighbor overlap integrals,
$a^{\ }_1=a^{\ }_2\equiv 1$ and $a^{\ }_3$ are the corresponding lattice constants, and
$\hat{c}^\dagger(\bs{k})=\left(\hat{c}_\uparrow^\dagger(\bs{k}),\hat{c}_\downarrow^\dagger(\bs{k})\right)$.
Here, 
$\hat{c}_s^\dagger(\bs{k}),\ s=\uparrow,\downarrow$,
creates an electron with momentum $\bs{k}$ and spin $s$,
while $\bs{\sigma}=(\sigma_1,\sigma_2,\sigma_3)$ denote the three Pauli matrices and $\sigma_0$ is the $2\times2$ unit matrix acting on spin space.
The real coefficient $\alpha$ parametrizes the strength of the Rashba spin-orbit coupling and 
\begin{equation}
\bs{g}(\bs{k}):=[-\sin\,( k^{\ }_2),\sin\, ( k^{\ }_1)]^{\mathsf{T}}.
\label{g}
\end{equation}
\end{subequations}

Within a mean-field treatment of the superconducting phase, the Bogoliubov-deGennes Hamiltonian 
with Eq.~\eqref{non-interacting H} as the non-interacting piece, written in terms of Nambu spinors
\begin{subequations}
\begin{equation}
\hat{\Psi}^\dagger(\bs{k})
:=\left[\hat{c}^\dagger_\uparrow(\bs{k}),\hat{c}^\dagger_\downarrow(\bs{k}),
\hat{c}^{\ }_\uparrow(-\bs{k}),\hat{c}^{\ }_\downarrow(-\bs{k})\right]
\end{equation}
is given by
\begin{equation}
\begin{split}
H^{\ }_{\mathrm{BdG}}:=&\,\sum_{\bs{k}\in\text{BZ}}
\hat{\Psi}^\dagger(\bs{k})
\begin{pmatrix}
\mathcal{H}(\bs{k})&\Delta(\bs{k})\\
\Delta^\dagger(\bs{k})&-\mathcal{H}^T(-\bs{k})
\end{pmatrix}
\hat{\Psi}(\bs{k}),
\label{BDG H}
\end{split}
\end{equation}
where
\begin{equation}
\Delta(\bs{k}):=
\left[d_0(\bs{k})\,\sigma_0+\bs{d}(\bs{k})\cdot \bs{\sigma}\right]
\mathrm{i}\sigma_2
\end{equation}
is the $2\times 2$ superconducting gap function decomposed in a
scalar singlet part 
\begin{equation}
d_0(\bs{k})=d_0(-\bs{k})
\label{s-symm}
\end{equation}
and a vector triplet part 
\begin{equation}
\bs{d}(\bs{k})=-\bs{d}(-\bs{k}).
\label{p-symm}
\end{equation}
The mixing of even- and odd-parity pairing is allowed due to the broken inversion 
symmetry~\cite{Gorkov01}. 
For simplicity, we assume $s$-wave pairing in the even-parity channel, that is,
$d_0(\bs{k})=\Delta_{\mathrm{s}}$ with  $\Delta_{\mathrm{s}}$ independent of $\bs{k}$.
The odd-parity component has ''$p$-wave'' character and 
has the generic form 
\begin{equation}
\bs{d}(\bs{k})= \Delta_{\mathrm{t}}\bs{g}(\bs{k}),
\label{d}
\end{equation}
resembling the Rashba term in the Hamiltonian (\ref{non-interacting H})~\cite{Sigrist09}.
Note that the quasiparticle spectrum is fully gapped in 2D,
while it can have line nodes in 3D.
\end{subequations}

Hamiltonian~\eqref{BDG H} has particle-hole symmetry and is invariant under time reversal. 
At the same time, the SU(2) spin-rotation symmetry is broken by the Rashba 
spin-orbit coupling due to the lack of inversion symmetry, for $x^{\ }_3 \to - x^{\ }_3 $.
As a consequence, it is not possible to define conserved spin currents.

\begin{figure}
\includegraphics[angle=0,scale=0.87]{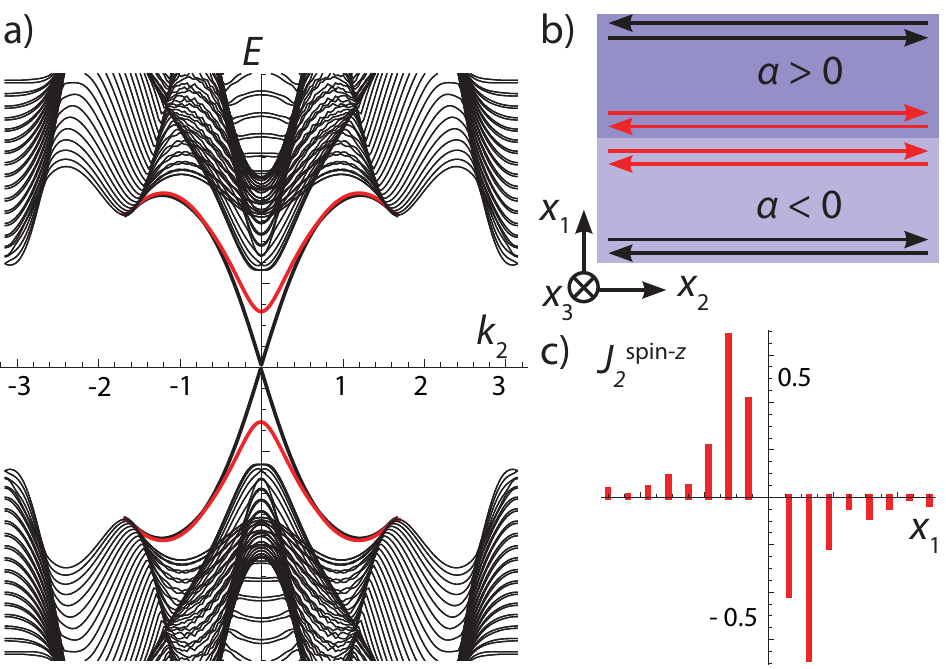}
\caption{\label{fig: boundary spectrum}
(Color online)
a)
Low energy spectrum of Hamiltonian~\eqref{BDG H} in a 2D strip geometry
with $t^{\ }_2/t^{\ }_1=1$, 
$\alpha/t^{\ }_1=0.7$, 
$\mu/t^{\ }_1=0.2$, 
$\mu_0/t^{\ }_1=2.2$,
$\Delta_{\mathrm{t}}/t^{\ }_1=0.5$, and
$\Delta_{\mathrm{s}}/t^{\ }_1=0.14$
in the twin boundary configuration~\eqref{sign-change alpha} as shown in b).
The highlighted midgap states are channels localized at the twin boundary, 
the other branch of midgap states are localized at the sample boundary.
c)
The spatial distribution of the spin current carried by the midgap modes,
$J^{\text{spin-}z}_2=\langle k^{\ }_2\, \sigma^{\ }_3\rangle$,
close to the twin boundary.
The spin-$z$ polarized current is an odd function of $x^{\ }_1$.
        }
\end{figure}

We illustrate here some basic properties of a twin boundary by diagonalizing the Hamiltonian~\eqref{BDG H}
in a strip geometry being of finite width along the $x^{\ }_1$- and infinitely long along the $x^{\ }_2$-direction. 
We consider a twin boundary located at $x^{\ }_1=0$ 
that is oriented with its normal parallel to the $x^{\ }_1$-direction. 
In the microscopic Hamiltonian, this is implemented by introducing
\begin{equation}
\begin{split}
\alpha &\to \alpha\, \mathrm{sgn}\, x^{\ }_1,\\
\mu &\to \mu+\mu_0\,\delta_{x^{\ }_1,0},\\
\Delta_{\mathrm{s}} &\to \Delta_{\mathrm{s}}\, \mathrm{sgn}\, x^{\ }_1,
\label{sign-change alpha}
\end{split}
\end{equation} 
where $\mu_0$ corresponds to a potential barrier at the twin boundary.
In this geometry, helical edge modes appear for topological reasons.
If the ground state of Hamiltonian~\eqref{BDG H} preserves TRS
and we consider the 2D case, where the ground state is fully gapped in the bulk,
Hamiltonian~\eqref{BDG H} belongs to symmetry class DIII in the classification of Ref.~\onlinecite{Schnyder08}.
In 2D, this symmetry class is equipped with a $\mathbb{Z}^{\ }_2$ topological attribute.
A topologically trivial state is realized for 
$|\Delta_{\mathrm{s}}|>|\Delta_{\mathrm{t}}|$,
while a nontrivial configuration with helical edge modes corresponds to
$|\Delta_{\mathrm{t}}|>|\Delta_{\mathrm{s}}|$.~\cite{Tanaka09,Santos10}
As a consequence, two completely decoupled twin domains would support two Kramers pairs of gapless
modes at their interface, where each pair represents the topological edge states of either twin domain.
Upon coupling the two twin domains, the gapless nature of the two Kramers pairs 
is lifted while preserving TRS, as evidenced by the gap that separates the pair of high-lighted bound states
in Fig.~\ref{fig: boundary spectrum}(a). Despite the appearance of this gap, these
Andreev bound states carry a spin supercurrent analogous to the edge states. 
The spin current has $x_3$ as quantization axis and is oriented opposite on the two sides of the 
twin boundary, as seen in Fig.~\ref{fig: boundary spectrum}(c). 
We should emphasize that these results are not a selfconsistent mean field solution to the problem
and as such cannot capture the modifications to the pairing potential due to the twin boundary.
The selfconsistent mean-field solution 
with a phenomenological spin-fluctuation based pairing interaction has been studied in Ref.~\onlinecite{Arahata12} 
and indeed shows that
\begin{itemize}
\item
the twin boundary supports helical Andreev bound states
that carry a (non-conserved) $z$-axis polarized spin current,
\item
for a certain range of singlet to triplet ratio $|\Delta_{\mathrm{s}}/\Delta_{\mathrm{t}}| \sim 1 $
the pairing state breaks TRS locally at the twin boundary, and
\item
after a secondary transition within the TRS breaking phase a finite magnetization
and a supercurrent emerge locally near the twin boundary.
\end{itemize}

It should be noted that the appearance of spin currents is not restricted
to the topological phase. 
Spin currents have also been predicted in the non-topological 
phase of noncentrosymmetric superconductors~\cite{Yip10}. 
Thus, spin currents are present for all ratios of $ |\Delta_{\mathrm{s}}  | $ and
$ |\Delta_{\mathrm{t}}  | $ on both sides of the topological transition. Ref.~\onlinecite{Arahata12} 
shows that the secondary phase transition is actually connected with the topological transition. 

In the next section, we are going to show how these results are
obtained within a GL 
description of noncentrosymmetric superconductors,
that is exclusively based on symmetry considerations.

\medskip
\section{
Ginzburg-Landau theory}
\label{sec: GL}

The aim of this section is to construct the GL free energy functional that can be used to describe a twin boundary in a noncentrosymmetric superconductor. Our analysis shall be valid both for a 3D system with tetragonal symmetry (where we identify the $c$-axis with the $x_3$-coordinate) and for a purely 2D system. We will thus impose translational symmetry for the 3D system in the $x_3$-direction and any terms containing the derivative in this direction will be neglected in the free energy expansion. In this case, we can always choose a gauge 
in which the 3-component of the electromagnetic vector potential vanishes identically $A^{\ }_3\equiv0$.

The class of noncentrosymmetric crystal structures discussed here obey tetragonal point group $ C_{4v} $, which is not invariant under reflection
at the basal plane. For our purpose we take, however, a step back starting from the centrosymmetric point group $ D_{4h} $. Introducing an inversion symmetry breaking ``auxiliary order parameter'' $\xi$ will be most convenient to implement the change between the two domains which are characterized by the opposite sign of $ \xi $. The order parameter $ \xi $ is real and belongs to the irreducible representation $\Gamma^{-}_2$ in $ D_{4h} $, and, thus, reduces the symmetry to $ C_{4v} $. This representation corresponds to symmetry introduced by Rashba spin-orbit coupling [$ \bs{g}(\bs{k}) \propto (-\sin k_2, \sin k_1)^{\mathsf{T}} $ belongs to $\Gamma^{-}_2$, see Tab.~\ref{tab: irrep}]. The superconducting order parameter has mixed-parity nature which we generate by an even- and an odd-parity component, $ \eta_s $ of the $\Gamma^{+}_1$- and $ \eta_p $ of the $\Gamma^{-}_2$-representation, respectively.

\begin{table}[t]
\caption{Relevant basis gap functions of 
one-dimensional irreducible representations 
for the tetragonal lattice symmetry $D^{\ }_{4h}$ (from Ref.~\onlinecite{Sigrist91b}).}
\centering 
\begin{tabular}{cl} 
\hline
Irreducible representation$\quad$ & Basis function \\
\hline 
$\Gamma^{-}_1$ & $\bs{d}=(\sin\,k^{\ }_1,\sin\,k^{\ }_2)^{\mathsf{T}}$\\
$\Gamma^{-}_2$ & $\bs{d}=(\sin\,k^{\ }_2,-\sin\,k^{\ }_1)^{\mathsf{T}}$\\
$\Gamma^{-}_3$ & $\bs{d}=(\sin\,k^{\ }_1,-\sin\,k^{\ }_2)^{\mathsf{T}}$\\
$\Gamma^{-}_4$ & $\bs{d}=(\sin\,k^{\ }_2,\sin\,k^{\ }_1)^{\mathsf{T}}$\\
\hline 
\end{tabular}
\label{tab: irrep}
\end{table}

\subsection{Bulk free energy}

The GL free energy functional is a scalar (fully symmetric under $ D_{4h} $) combination of the order parameters $ \eta_s $, $ \eta_p $ and $ \xi $ and the covariant gradients $ \bs{D} = \bs{\nabla} - i \bs{A} $, where we choose the units so that $ 2 \pi /\Phi_0 =1 $ ($\Phi_0 = hc/2e $ the flux quantum). Note that $ \xi $ will be taken here as a constant system parameter reducing the crystal symmetry. 
The free energy can be written as
\begin{subequations}
\begin{equation}
F[\eta^{\ }_s,\eta^{\ }_p,\bs{A}]
:=F^{\ }_0
+F^{\ }_{\mathrm{M}},
\end{equation}
where the magnetic field energy is given by
\begin{equation}
F^{\ }_{\mathrm{M}}[\bs{A}]
:=
\frac{1}{8\pi}
\int \mathrm{d}^d\bs{x}\,
(\bs{\nabla}\times\bs{A})^2
,
\end{equation}
and the superconducting free energy $F^{\ }_0$ contains all symmetry-allowed terms 
up to fourth order in $\eta_{s,p}$ as well as the gradient terms up to second order in 
$\eta_{s,p}$ and $\bs{D}$. We decompose $F^{\ }_0$ into three parts
\begin{equation}
F^{\ }_0[\eta^{\ }_s,\eta^{\ }_p,\bs{A}]
:=\int \mathrm{d}^d\bs{x}
\left(f^{\ }_s+f^{\ }_p+f^{\ }_{sp}
\right),
\end{equation}
where 
\begin{equation}
f^{\ }_\ell
:=
 a^{\ }_{\ell} \vert \eta^{\ }_{\ell}\vert^{2}
+b^{\ }_{\ell} \vert \eta^{\ }_{\ell} \vert^{4} 
+\gamma^{(0)}_{\ell} \vert \bs{D} \eta^{\ }_{\ell}\vert^{2},
\qquad
\ell=s,p,
\end{equation}
and
\begin{equation}
\begin{split}
f^{\ }_{sp}
:=&\,
d\, \xi \left(  \eta_{s}^{\ast} \eta^{\ }_{p} + \eta^{\ }_{s}\eta_{p}^{\ast} \right) 
+c^{\ }_{1} \vert\eta^{\ }_{s}\vert^{2} \vert \eta^{\ }_{p} \vert^{2} 
+ c^{\ }_{2} \left( \eta_{s}^{\ast 2} \eta_{p}^{2} + \eta_{s}^{2} \eta_{p}^{\ast 2} \right) 
\\
&\,
+ \gamma^{(0)}_{0}\, \xi \left[  \left(  \bs{D} \eta^{\ }_{s} \right)^{\ast}\cdot  \bs{D} \eta^{\ }_{p} 
+ \bs{D}\eta^{\ }_{s} \cdot \left( \bs{D}\eta^{\ }_{p} \right)^{\ast} \right].
\end{split}
\label{f0}
\end{equation}
\label{bulk free energy}
\end{subequations}
The following phenomenological parameters appear: 
$a^{\ }_{\ell},\ b^{\ }_{\ell},\ \gamma^{(0)}_{\ell}$, $\ell=s,p$, as well as
$d,\ c^{\ }_{1}, c^{\ }_{2}$ and $\gamma^{(0)}_{0}$. 
All these parameters are taken independent of temperature $T$ except for
$a^{\ }_{\ell}=a'_{\ell}(T-T^{\ }_{\mathrm{c},\ell})$, $\ell=s,p$, with
$T^{\ }_{\mathrm{c},\ell}$ being the bulk
transition temperature of the ficticious homogeneous superconductor of 
$s$- or $p$-wave type in the absence of parity mixing.

The free energy is constructed to be invariant under the TRS $\mathcal{T}$
\begin{subequations}
\begin{eqnarray}
\eta&\stackrel{\mathcal{T}}{\longrightarrow}& \eta^*,
\label{TRS1a}
\\
\bs{A} &\stackrel{\mathcal{T}}{\longrightarrow}& -\bs{A},
\label{TRS1b}
\end{eqnarray}
\label{TRS1}
\end{subequations}
and under the local U(1) gauge transformation $\Phi$ 
\begin{equation}
\begin{split}
\eta\stackrel{\Phi}{\longrightarrow} e^{\mathrm{i} \varphi(\bs{x})} \eta,
\qquad
\bs{A} \stackrel{\Phi}{\longrightarrow} \bs{A} +\bs{\nabla} \varphi(\bs{x}).
\end{split}
\end{equation}
Here, $\bs{A}=(A^{\ }_1,A^{\ }_2)$ are the $x^{\ }_1$ and $x^{\ }_2$ 
components of the electromagnetic vector potential.

\begin{subequations}
We now consider the homogeneous bulk superconducting phase. The onset of superconductivity is 
determined by the second-order terms. For the centrosymmetric system, $ \xi = 0 $, either the
even- ($\eta_s$) or the odd-parity ($\eta_p $) appears at the phase transition, depending on which has
the highest $ T_c $. The noncentrosymmetric situation with $ \xi \neq 0 $ yields a mixture
of both. Thus, the resulting state is TRS whereby both, $\eta^{\ }_s$ and $\eta^{\ }_p$
are non-vanishing and their relative phase
\begin{equation}
\phi:=\left(\mathrm{arg}\,\eta^{\ }_s - \mathrm{arg}\,\eta^{\ }_p\right)\ \mathrm{mod}\ 2\pi,
\end{equation}
is a well-defined gauge-invariant quantity that
changes sign under time reversal.
Thus, only the configurations 
\begin{equation}
\phi=0,\pi,
\label{cases phi}
\end{equation}
are compatible with TRS.
Observe that these are the states favored by the lowest order coupling term in Eq.~\eqref{f0}, which is rewritten as 
\begin{equation}
2d\xi |\eta^{\ }_s\eta^{\ }_p| \cos \phi.
\label{quadratic parity mixing}
\end{equation}
\end{subequations}

Before turning to the case with twin boundary, we note that the bulk free energy~\eqref{bulk free energy} is also invariant under the spatial inversions $P_1$ and $P_2$ of the $x_1$ and $x_2$ coordinates, respectively,
\begin{equation}
x_i\stackrel{P_i}{\longrightarrow} -x_i,\qquad i=1,2.
\label{inversion}
\end{equation}
Modeling the twin boundary at $x_1=0$, say, is now conveniently achieved by endowing $\xi$ with the spatial dependence
\begin{equation}
\xi(x_1)=
\begin{cases}
+1& x_1>0,\\
-1& x_1<0.
\end{cases}
\label{xi spatial dependence}
\end{equation}
Thus, the twin boundary explicitly breaks the inversion symmetry $P_1$, while $P_2$ stays intact. 
The relevant symmetry group of our model of a twin boundary in a noncentrosymmetric superconductor defined by 
Eqs.~\eqref{bulk free energy} and \eqref{xi spatial dependence}, including $ \Phi, \mathcal{T} $ and $ P_2 $ is thus
\begin{equation}
{\cal G} = \mathrm{U}(1)\times\mathbb{Z}_2\times\mathbb{Z}_2.
\end{equation}

\subsection{TRS breaking superconductivity at the twin boundary}

Having clarified some aspects of the bulk superconducting states, 
let us now consider the system with twin boundary as defined by Eq.~\eqref{xi spatial dependence}.
In view of Eq.~\eqref{quadratic parity mixing}, one concludes that 
the relative phase $\phi$ changes by $ \pm \pi$ from one twin domain to the other.

This change in phase can be accounted for in two ways:
(i) If one of the two order parameter components, $\eta^{\ }_s$ or $\eta^{\ }_p$, changes sign and vanishes at the twin boundary,
such that $\phi$ changes abruptly by $\pi$ or 
(ii) if both $\eta^{\ }_s$ and $\eta^{\ }_p$ do not vanish by turning complex at the twin boundary,
such that $\phi$ changes continuously across the boundary.
Here, (i) preserves the TRS, while in (ii) TRS is broken 
near the twin boundary by violation of Eq.~\eqref{cases phi}.

If one of the two order parameters is clearly dominant in the bulk, 
case (i) is realized~\cite{Iniotakis08,Arahata12}.
The dominant order parameter remains phase-coherent while the subdominant one changes sign at the twin boundary.
Due to the TRS, the supercurrent 
vanishes along the twin boundary.
In contrast, if $\eta^{\ }_s$ and $\eta^{\ }_p$ are nearly degenerate, the spontaneous TRS breaking 
(ii) is energetically favorable.
In this state, the relative phase of the superconducting order parameters changing continuously assumes one 
of two degenerate values $\phi=\pm\pi/2$ at the twin boundary, reflecting the $\mathbb{Z}_2$ nature of the symmetry breaking.  
The inversion symmetry $P_2$ is still intact in this state. 
It follows that, in spite of the broken TRS, the supercurrent along the twin boundary, 
which is associated with $P_2$, remains zero, as we will discuss below.

\subsection{Spontaneous magnetization and currents}

We have seen that the superconducting order near the twin boundary can break TRS spontaneously.
This triggers the question whether the TRS breaking superconducting state can drive a spontaneous magnetization
and supercurrent near the twin boundary.
To answer this, we extend the  
free energy~\eqref{bulk free energy} by adding terms that stem from
the magnetization $\bs{m}=(m^{\ }_1,m^{\ }_2)$ and $m^{\ }_3$ as a 3-component order parameter
\begin{subequations}
\begin{equation}
\begin{split}
F[\eta^{\ }_s,\eta^{\ }_p,\bs{m},m^{\ }_3,\bs{A}]
:=
&\,
F^{\ }_0
+
F^{\ }_\mathrm{M}
\\
&\,+
\int \mathrm{d}^d\bs{x}
\left(f^{\ }_{\mathrm{m}}+f^{\ }_{\text{hel}}+f^{\ }_{\text{s-H}}
\right),
\end{split}
\label{free energy with magnetic coupling}
\end{equation}
where $F^{\ }_0$ and $F^{\ }_\mathrm{M}$ are defined in Eq.~\eqref{bulk free energy} while
\begin{equation}
f^{\ }_{\mathrm{m}}
:=
\sum_{i,j=1}^3
(\bs{\nabla}m^{\ }_i)^{\mathsf{T}}\bs{\Xi}^{(ij)}(\bs{\nabla}m^{\ }_j)
+\sum_{i=1}^3\chi^{-1}_{ii} m^{2}_i + b^{\ }_3 m^{4}_3,
\end{equation}
is the free energy for the magnetic order parameter
and
\begin{equation}
\begin{split}
f^{\ }_{\text{hel}}
:=
&\,
\mathrm{i} \gamma^{(1)}_0 (\bs{e}^{\ }_3\times \bs{m})\cdot
\left[
\left(
\eta^{*}_{s} \bs{D}\eta^{\ }_{p} 
+
\eta^{*}_{p} \bs{D}\eta^{\ }_{s} 
\right)
-\mathrm{c.c.}\right]
\\
&\,
+\mathrm{i} \sum_{\ell=s,p} 
\gamma^{(1)}_\ell \xi (\bs{e}^{\ }_3\times \bs{m})\cdot
\eta^{*}_{\ell}\bs{D}\eta^{\ }_{\ell}
,
\end{split}
\label{f hel}
\end{equation}
\begin{equation}
\begin{split}
f^{\ }_{\text{s-H}}
:=&\,
\mathrm{i} \gamma^{(3)}_0 \xi m^{\ }_3
\left[
\left(  \bs{D} \eta^{\ }_{s} \right)^{\ast} \times \bs{D}\eta^{\ }_{p} 
-\mathrm{c.c.}\right]
\\
&\,
+
\mathrm{i} \sum_{\ell=s,p} 
\gamma^{(3)}_\ell m^{\ }_3
\left(  \bs{D} \eta^{\ }_{\ell} \right)^{\ast} \times \bs{D}\eta^{\ }_{\ell} 
,
\end{split}
\label{f s-h}
\end{equation}
are the helical and spin-Hall type coupling terms to the superconducting order parameters, respectively, which are obtained
as the only scalar combinations to this order including these order parameters. 
In a microscopic theory, they correspond to terms induced by spin-orbit coupling, rendering the electrons spin a non-conserved quantity.
\label{magnetic free energy}
\end{subequations}

Here, $\chi^{\ }_{11}=\chi^{\ }_{22}$ and $\chi^{\ }_{33}$ are the components of the diagonal static uniform magnetic susceptibility tensor. The paramagnetic nature of the bulk state requires 
\begin{equation}
\chi^{\ }_{11}=\chi^{\ }_{22}>0,
\qquad
\chi^{\ }_{33}>0.
\end{equation}
The components of the positive definite $2\times2$ stiffness tensor $\bs{\Xi}^{(ij)},\ i,j=1,\cdots,3$ are related to the static susceptibility $\chi_{ij}(\bs{q})$ via
\begin{equation}
\Xi^{(ij)}_{\mu,\nu}
:=\lim_{\bs{q}\to 0}\, \partial_{q^{\ }_\mu} \partial_{q^{\ }_\nu} \chi^{-1}_{ij}(\bs{q}),
\qquad
\mu, \nu\in\{1,2\}
.
\end{equation}
assuming that $ \chi_{ij}(\bs{q}) $ is maximal at $ \bs{q} = 0 $. 
where the tetragonal symmetry requires the vanishing of 
$\bs{\Xi}^{(13)}=\bs{\Xi}^{(31)}=0$
and
$\bs{\Xi}^{(23)}=\bs{\Xi}^{(32)}=0$.
Further, the parameter $b^{\ }_3>0$ and we introduced
$\gamma^{(i)}_0$, $\gamma^{(i)}_\ell$, $i=1,3$, $\ell=s,p$,
to parametrize the symmetry allowed linear coupling terms of magnetization and superconductivity.

As discussed in Sec.~\ref{sec: microscopics}, we expect the appearance of spin-carrying modes at the twin boundary,
both in the TRS and time-reversal broken phase. 
To describe spin currents within the GL formalism, however, one has to go beyond the set of order parameters introduced here. More precisely, one has to consider the admixture of superconducting order parameters that are not allowed in the bulk, but emerge at the (twin) boundary due to the translational symmetry breaking.
For example, consider the admixture of a triplet order parameter $\tilde{\eta}_p$  in the irreducible representation $\Gamma^-_1$ in addition to $\eta_p$  in the irreducible representation $\Gamma^-_2$.
A spin current corresponds to  a phase winding in the order parameter for upspins
\begin{equation}
\Delta_{\uparrow\uparrow}(\bs{k},\bs{r})
\sim
(k_2+\mathrm{i} k_1)[\eta_p(\bs{r})-\mathrm{i}\tilde{\eta}_p(\bs{r})]
\propto
e^{-\mathrm{i}\bs{q}\cdot\bs{r}},
\end{equation}
while the order parameter for downspins
\begin{equation}
\Delta_{\downarrow\downarrow}(\bs{k},\bs{r})
\sim
(k_2-\mathrm{i} k_1)[\eta_p(\bs{r})+\mathrm{i}\tilde{\eta}_p(\bs{r})]
\propto
e^{+\mathrm{i}\bs{q}\cdot\bs{r}}
\end{equation}
has a phase winding in the opposite direction. Such a spin current is realized when $(\eta_p,\tilde{\eta}_p)\propto(\cos\,\bs{q}\cdot\bs{r},\sin\,\bs{q}\cdot\bs{r})$, but cannot be described when only one order parameter component is present.
As the spin currents are omnipresent and not related to any phase transitions in our model, we refrain from considering them further in the interest of simplicity.

The superconducting charge current, in contrast, is defined as usual
\begin{equation}
\bs{J}:= \frac{\delta F}{\delta \bs{A}}.
\label{supercurrent}
\end{equation}
Considering the symmetry transformations of the spin current and the supercurrent, we note that the spin current preserves both $P_2$ and $\mathcal{T}$, while a nonzero supercurrent along the twin boundary breaks both $P_2$ and $\mathcal{T}$.

\subsection{Helical contribution}

We discuss now in more detail
the helical coupling to $\bs{m}$ in linear order, Eq.~\eqref{f hel}, which 
is specific to noncentrosymmetric systems. For a homogeneous inplane magnetization, it enforces a gradient on
the phase of the superconducting order parameter~\cite{ED89,DIM03,KAUR05,AG07,Neupert11},
because $\bs{m}$ is coupled to a supercurrent term of the form
\begin{equation}
\eta^* \bs{D}\eta
-
\eta (\bs{D}\eta)^*.
\end{equation}
In fact, if the relation
\begin{equation}
(\gamma^{(1)}_0,\gamma^{(1)}_s,\gamma^{(1)}_p)
=\alpha^{\ }_{10}
(\gamma^{(0)}_0,\gamma^{(0)}_s,\gamma^{(0)}_p),
\end{equation}
is valid for a real number $\alpha^{\ }_{10}$,
we can cast the Ginzburg-Landau equations for $\bs{m}$ in the form
\begin{equation}
\sum_{j=1}^2
\sum_{\mu,\nu=1}^2
\Xi^{(ij)}_{\mu,\nu}
\partial^{\ }_\mu
\partial^{\ }_\nu
\, m^{\ }_j
=
-\frac{
\alpha^{\ }_{10}}
{2}
\xi 
(\bs{e}^{\ }_3\times \bs{J}^{(0)})^{\ }_i
+
\chi^{-1}_{ii} m^{\ }_i
,
\label{GL-mx-my}
\end{equation}
for $i=1,2$,
where $\bs{J}^{(0)}$ is the supercurrent defined in Eq.~\eqref{supercurrent} with $\bs{m}=0$ and $m_3=0$.
The form of Eq.~\eqref{GL-mx-my} makes evident that the supercurrent acts as a source term for
$\bs{m}$.
Hence, if $\bs{J}^{(0)}$ vanishes, 
Eq.~\eqref{GL-mx-my} will only have the trivial solution $\bs{m}\equiv0$ owing to the positive definiteness of
$\chi^{-1}$ and $\bs{\Xi}^{(ij)}$.
Conversely, a finite supercurrent, which necessitates broken TRS, inevitably drives $\bs{m}$ to a finite value, oriented perpendicular
to both $ \bs{e}^{\ }_3 $ and $\bs{J}^{(0)} $.

\begin{figure}
 \includegraphics[angle=0,scale=0.55,page=1]{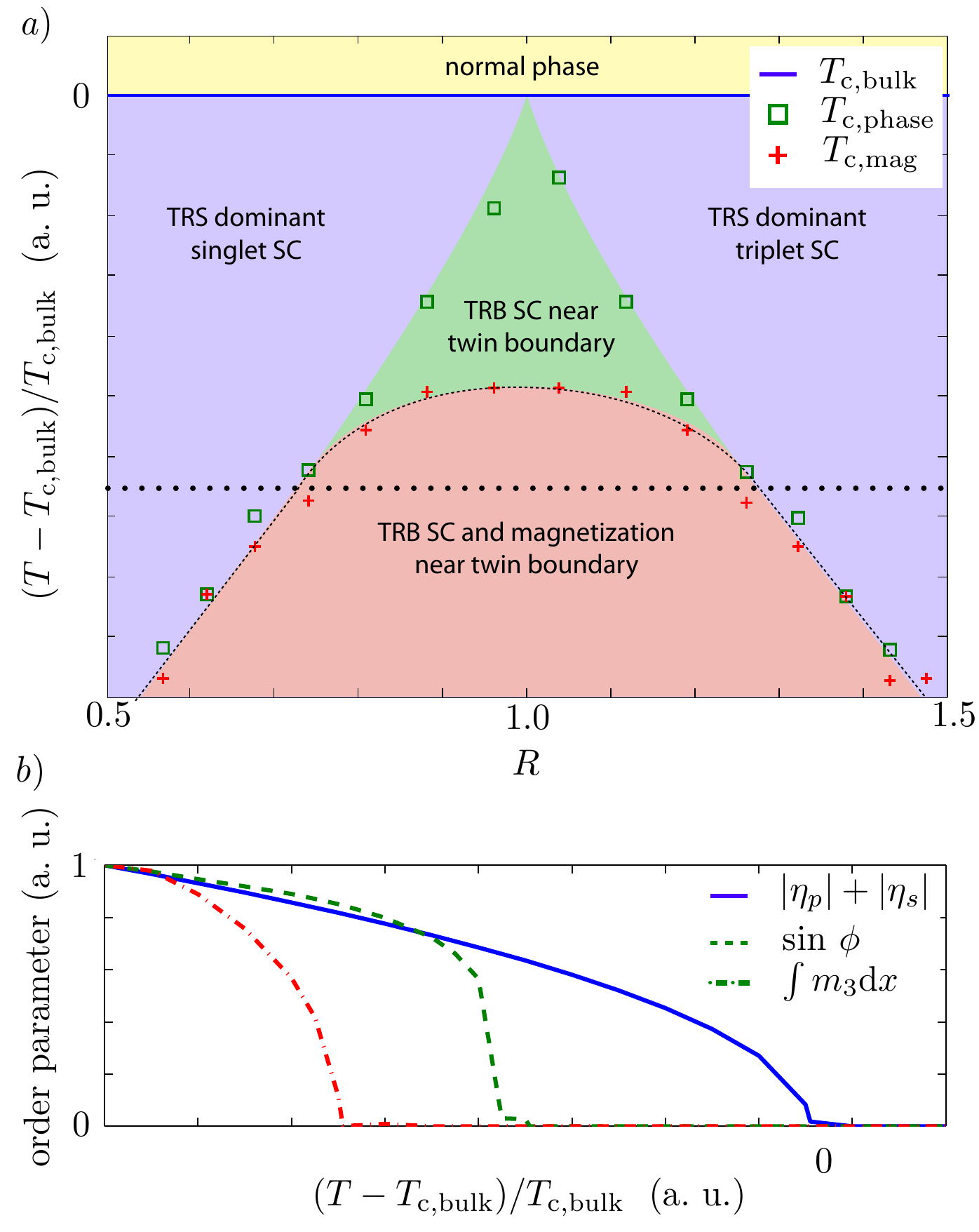}
\caption{\label{fig: phase_diagram}
(Color online)
a) Phase diagram of superconducting and magnetic order at the twin boundary
as a function of temperature,
where $T^{\ }_{\mathrm{c,bulk}}$ is the bulk transition temperature,
and parity mixing parameter $R:=2|\eta^{\ }_p/\eta^{\ }_s|/(|\eta^{\ }_p/\eta^{\ }_s|+1)$.
In the TRS phases, the relative phase of $\eta^{\ }_p$ and $\eta^{\ }_s$ is $\phi=0,\pi$ everywhere, 
while in the time-reversal broken phases, $\phi\neq0,\pi$ 
near the twin boundary.
This time-reversal breaking (TRB) by the superconducting state promotes a
ferromagnetic order $m^{\ }_3$ near the twin boundary at sufficiently low temperatures.
b)
Cut through the phase-diagram at $R=0.88$ where the order parameters 
$|\eta^{\ }_s|+|\eta^{\ }_p|$, $\sin\phi$ at the mesh point next to the twin boundary,
and
$\int m^{\ }_3 \mathrm{d}x^{\ }_1$ are chosen to signal the phase transitions
towards the superconducting phase, the time-reversal breaking phase and the magnetized phase
via the continuous, dashed, and dotted-dashed line, respectively.
The results were obtained for a discretized mesh with 300 sites in $x^{\ }_1$-direction~\cite{footnote1}.
        }
\end{figure}

\subsection{Spin Hall contribution}

Let us now turn to the spin-Hall coupling involving the out-of-plane magnetization $m^{\ }_3$, Eq.~\eqref{f s-h}, which 
is generically present also in centrosymmetric tetragonal superconductors.
We will find that via a second order phase transition $m^{\ }_3$ and a supercurrent~\eqref{supercurrent} 
appears \emph{simultaneously}, which necessitates a breaking of both TRS $\mathcal{T}$ and inversion $P_2$. 
This is different from the case of the in-plane magnetization $\bs{m}$, which is driven by an \emph{existing} supercurrent via Eq.~\eqref{GL-mx-my}.
To illustrate the second order phase transition of $m^{\ }_3$ and the supercurrent, we consider the geometry with twin boundary and impose two gauge-fixing conditions:
(i) We shall assume that there is no current flowing across the twin boundary (in $x^{\ }_1$ direction), so that we can choose a gauge in which $A^{\ }_1\equiv0$. 
(ii) We impose translation invariance along the twin boundary (in $x^{\ }_2$ direction), and choose the gauge such that there is no spatial dependence along the direction $ \bs{e}_2 $. 
Then, it is convenient to combine $m^{\ }_3$ and $A^{\ }_2$ 
into a new real two-component order parameter
\begin{subequations}
\begin{equation}
G:=(m^{\ }_3,\xi A^{\ }_2)^{\mathsf{T}}.
\end{equation}
To second order in $G$, the corresponding free energy density is rewritten as
\begin{equation}
\begin{split}
f^{\ }_G=&\,
(\partial^{\ }_1G)^{\mathsf{T}}
\begin{pmatrix}
\Xi^{(33)}_{1,1}
&
0
\\
0
&
(8\pi)^{-1}
\end{pmatrix}
(\partial^{\ }_1G)
+
G^{\mathsf{T}}
M(\eta^{\ }_{s},\eta^{\ }_{p})
G,
\label{FG1}
\end{split}
\end{equation}
where 
\begin{equation}
M(\eta^{\ }_{s},\eta^{\ }_{p}):=\begin{pmatrix}
\chi^{-1}_{33}
&
\xi \partial^{\ }_1 h^{(3)}/2
\\
\xi \partial^{\ }_1 h^{(3)}/2
&
h^{(0)}
\end{pmatrix},
\label{M}
\end{equation}
and we defined for $i=0,1,3$
\begin{equation}
h^{(i)}(\eta^{\ }_{s},\eta^{\ }_{p}):=
\gamma^{(i)}_s|\eta^{\ }_{s}|^2+\gamma^{(i)}_p|\eta^{\ }_{p}|^2+
2\gamma^{(i)}_0\xi |\eta^{\ }_{s}||\eta^{\ }_{p}|\cos\phi,
\end{equation}
\end{subequations}
all of which are even functions of $x^{\ }_1$ and $h^{(0)}>0$.
In particular, the matrix $M(\eta^{\ }_{s},\eta^{\ }_{p})$ is an even function of $x_1$.
Observe that the free energy~\eqref{FG1} has a $\mathbb{Z}^{\ }_2$ symmetry defined by
\begin{equation}
G\rightarrow -G,
\label{Z2 symmetry}
\end{equation}
which is nothing but the representation of the inversion symmetry $P_2$ defined in Eq.~\eqref{inversion}.

The GL equation that results form the free energy~\eqref{FG1} resembles a Schr\"odinger equation
for a spinor-valued field, with the matrix $M$ playing the role of a potential.
The matrix $M$ is 
asymptotically positive definite for $x^{\ }_1\to \pm \infty$. 
We seek the solution $G$ that satisfies the boundary conditions $|G|\to 0$ as $x^{\ }_1\to \pm \infty$
and is lowest in energy, i.e., the lowest bound state of this potential.
Since the matrix $M$ is an even function of $x_1$,
we can infer that the lowest energy bound state, if it exists, is even in $x_1$ as well and has no nodes.
As a corollary, a nontrivial solution for $m_3$ will be an even function of $x_1$ while 
$A_2$ and the supercurrent along the twin boundary will be odd functions of $x_1$ 
with one node at $x_1=0$.
The existence of a non-trivial solution requires $M$ to be not positive definite in some region near the twin boundary,
while only the trivial solution $G\equiv0$ exists, if $M$ is positive definite everywhere.
We will now argue that these two situations are realized in two limiting configurations for $\eta^{\ }_s$ and $\eta^{\ }_p$.

On one hand, if one of the superconducting order parameters is dominant, say $|\eta^{\ }_{s}|\gg|\eta^{\ }_{p}|$,
then $|\partial^{\ }_1\eta^{\ }_{s}|$ is small, since the dominant order parameter remains coherent across the twin boundary, but also
$|\partial^{\ }_1\eta^{\ }_{p}|$ is small by the very smallness of $|\eta^{\ }_{p}|$. It follows that the off-diagonal elements of $M$ in Eq.~\eqref{M} will be small. 
Since the diagonal terms of $M$ are positive, $M$ is (nearly) positive definite in this limit. Hence, with the boundary conditions in place, no spontaneous magnetization or current develops.

On the other hand, if $|\eta^{\ }_{s}|\approx|\eta^{\ }_{p}|$, such that we expect a TRS broken state at the twin boundary,
we can approximate the off-diagonal terms in $M$ as
\begin{equation}
\partial^{\ }_1 h^{(3)}/2
\approx
\gamma^{(3)}_0 |\eta^{\ }_{s}||\eta^{\ }_{p}|\partial^{\ }_1\, \cos\phi.
\label{partial h estimate}
\end{equation}
This term will turn $M$ into an indefinite matrix, whenever it is nonzero.
In fact, as $\phi$ changes smoothly from $0$ to $\pi$ across the twin boundary,
we expect this term not to be small in the vicinity of the boundary.
In this situation we anticipate nontrivial solutions $G\equiv G_0 (x_1) \neq0$, 
whereby $G_0$ and $-G_0$ are degenerate due to the inversion symmetry $P_2$, Eq.~\eqref{Z2 symmetry}, 
signaling the violation of this symmetry. 
Thus, the emergence of $G$ is a second spontaneous breaking of a $\mathbb{Z}^{\ }_2$ symmetry via a second-order phase transition in our model.

\begin{figure}
 \includegraphics[angle=0,scale=0.55,page=2]{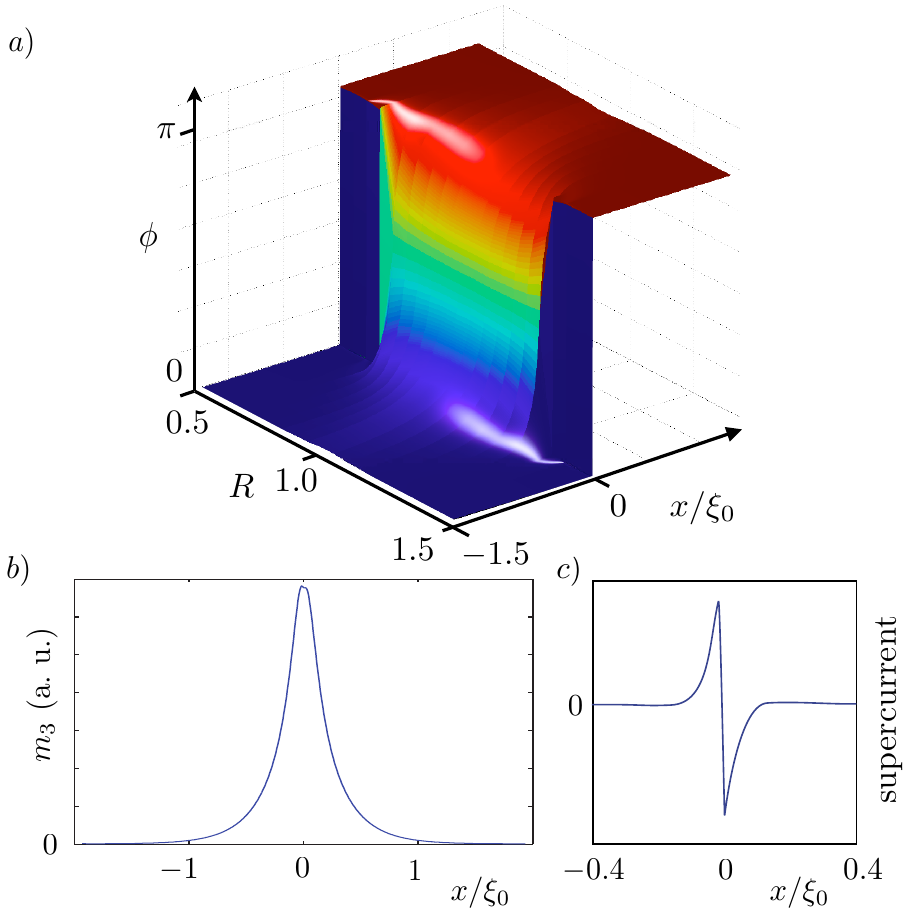}
\caption{\label{fig: phase current mag}
(Color online)
a) Spatial dependence of 
the relative phase $\phi$ between $\eta^{\ }_p$ and $\eta^{\ }_s$ 
along a cut at constant temperature denoted by the dotted line in Fig.~\ref{fig: phase_diagram}
.
b)
The magnetization $m^{\ }_3$ and 
c)
the supercurrent $J^{\ }_2$ as a function of $x^{\ }_1$  for 
$R:= 2|\eta^{\ }_p/\eta^{\ }_s|/(|\eta^{\ }_p/\eta^{\ }_s|+1)=1$
near the twin boundary in the TRS breaking phase with finite magnetization.
The position is measured in units of the mean coherence length
$\xi^{-2}_{0}:=(a^{\ }_s+a^{\ }_p)/(\gamma^{(0)}_s+\gamma^{(0)}_p)$.
These results are obtained for a discretized mesh with 3000 sites in $x^{\ }_1$-direction
and $a'_s=a'_p=0.01$
~\cite{footnote1}.
        }
\end{figure}
The physical implication of $G\neq 0$ near the twin boundary is two-fold:
First, the magnetization $m^{\ }_3$ is nonvanishing. 
Second, the finite gauge field $A^{\ }_2(x^{\ }_1)$ yields a supercurrent along the $ \bs{e}_2 $-direction, $J^{(0)}_2= A^{\ }_2 h^{(0)}$,
and
drives the magnetization $m^{\ }_1$ as described by Eq.~\eqref{GL-mx-my}. 
All these three order parameters $A^{\ }_2$, $m^{\ }_3$, and $m^{\ }_1$ enter the expression for the supercurrent parallel to the twin boundary that reads
\begin{equation}
\begin{split}
J^{\ }_2
:=&\,
J^{(0)}_2
+
\frac{\partial f^{\ }_{\text{s-m}}}{\partial A^{\ }_2}\\
=&\,
2 A^{\ }_2 h^{(0)}
-2 \xi
 m^{\ }_1 h^{(1)}
+
 m^{\ }_3 \partial^{\ }_1 h^{(3)}
.
\end{split}
\label{supercurrent2}
\end{equation}

The qualitative line of reasoning outlined here is confirmed by the results of the explicit numerical 
minimization of the GL free energy~\eqref{free energy with magnetic coupling}
that provide the phase diagram shown in Fig.~\ref{fig: phase_diagram}.
It resolves the consecutive breaking 
of the two $\mathbb{Z}_2$-symmetries $\mathcal{T}$ and $P_2$ 
with decreasing temperature as two distinct phase transitions.
The spatial profiles of the order parameters are shown in Fig.~\ref{fig: phase current mag}.
These results confirm the expectation that $m^{\ }_3(x^{\ }_1)$ is an even function of $x_1$,
while $A^{\ }_2(x^{\ }_1)$ is an odd function of $x^{\ }_1$.
Thus, the supercurrent $J^{\ }_2(x^{\ }_1)$ flows in opposite directions slightly to the left and to
the right of the twin boundary.

\medskip
\section{
Discussion and conclusion}

In this paper, we studied the sequence of symmetry-breaking transitions
at twin-boundaries in noncentrosymmetric superconductors.
We found that the superconducting order may spontaneously break TRS at the boundary
if singlet and triplet pairing is of comparable magnitude, corresponding to the regime where
a topological transition occurs.~\cite{Tanaka09,Arahata12}
This symmetry breaking is associated with a nontrivial relative phase between
singlet and triplet pairing order parameters. 
In the phase where TRS is broken,
the emergence of a supercurrent and a finite magnetization at the twin boundary
is favored. Both the supercurrent and the magnetization break spontaneously the 
inversion symmetry along the twin boundary, in addition to the already broken TRS.
Our model for the twin boundary thus shows two consecutive breakings of $\mathbb{Z}_2$ symmetries,
resulting in a total of $2\times 2=4$ degenerate low-symmetry states. 
One expects that these 4 states form domains
along the twin boundary and are separated by topological defects that can only exist on the 
twin boundary. The topological defect associated with the first $\mathbb{Z}_2$ symmetry (TRS) 
is a vortex that carries a fractional magnetic flux~\cite{Iniotakis08}, while the defect associated 
with the second $\mathbb{Z}_2$ symmetry (inversion) is a magnetic domain wall.
We leave the investigation of the interplay of these defects as a subject for future studies.

We observe that the spontaneous supercurrents and the spin magnetizations
are always coupled to one another through the spin current at the twin boundary.
This is not a feature exclusive to the topologically non-trivial phase, but occurs 
in a range of parameters on both sides of the topological transitions. The spin Hall nature of
this interrelation between spin-current, charge-current and the spin magnetization is the basis
of many devices in spintronics. 

On closing, we shall briefly relate to the experimental example of the orthorombic noncentrosymmetric
superconductor LaNiC$_2$. Muon spin rotation measurements showed that 
its superconducting state breaks TRS,~\cite{Hillier09} which raises 
the question about its pairing symmetry. Symmetry arguments show that the ``usual'' form
of a TRS breaking superconducting phase is impossible, due to the lack of a degenerate multi-component
order parameter. 
In analogy to our results for twin boundaries, one might speculate whether TRS
is not broken by the bulk pairing symmetry of the superconducting order parameter,
but via local TRS breaking bound states at defects~\cite{Sigrist91a} in this material. 
The fact that the TRS breaking phase only appears as subsequent transition below $ T_c $ is, however,
in contradiction with the experimental observation that the intrinsic magnetism appears at the onset
of superconductivity~\cite{Hillier09}. 

\medskip
\section*{Acknowledgments}

We would like to thank K. Aoyama, P. Brydon, S. Fujimoto, J. Goryo, C. Iniotakis, D. Manske, A.C. Mota, and A. Schnyder for helpful discussions.
This work was supported in part by the Swiss National Science Foundation, the NCCR MaNEP, and the Pauli Center for Theoretical Studies of the ETH Zurich. E.A. was supported by a Grant-in-Aid from JSPS.


\begin{thebibliography}{99}
\bibitem{footnote1}
These results were obtained from the free energy~\eqref{free energy with magnetic coupling},
with the parameter values
$b^{\ }_s=b^{\ }_p=c^{\ }_1=0.01$,
$d=0.03$, $c^{\ }_2=0.004$, 
$\gamma^{(0)}_s=\gamma^{(0)}_p=\gamma^{(3)}_s=\gamma^{(3)}_p=\gamma^{(3)}_0=0.05$,
$b^{\ }_3=\chi^{\ }_{33}=0.1$, $\gamma^{(0)}_0=0.04$,
$\gamma^{(1)}_0=0.003$, 
and 
$\gamma^{(1)}_s=\gamma^{(1)}_p=0.005$
.
With this, we have chosen all those parameters pairwise equal, that multiply the equivalent terms 
of the singlet and triplet order parameter. Deviations from this symmetric choice will shift the phase 
boundaries in the phase diagram, but its qualitative features are expected to survive.
\end{thebibliography}
\end{document}